\documentclass[sigconf]{acmart}
\usepackage{graphicx}
\usepackage{url}
\usepackage{tabularx}
\usepackage{caption}
\usepackage{subcaption}
\usepackage{lscape}
\usepackage{enumitem}

\AtBeginDocument{%
  \providecommand\BibTeX{{%
    \normalfont B\kern-0.5em{\scshape i\kern-0.25em b}\kern-0.8em\TeX}}}

\copyrightyear{2023} 
\acmYear{2023} 
\setcopyright{rightsretained} 
\acmConference[CHI EA '23]{Extended Abstracts of the 2023 CHI Conference on Human Factors in Computing Systems}{April 23--28, 2023}{Hamburg, Germany}
\acmBooktitle{Extended Abstracts of the 2023 CHI Conference on Human Factors in Computing Systems (CHI EA '23), April 23--28, 2023, Hamburg, Germany}%\acmDOI{10.1145/3544549.3585583}
\begin{document}

\title[Regret, Delete, (Do Not) Repeat]{Regret, Delete, (Do Not) Repeat: An Analysis of Self-Cleaning Practices on Twitter After the Outbreak of the COVID-19 Pandemic}

\author{Nicol\'{a}s E. D\'{i}az Ferreyra}
\authornote{Both authors contributed equally to this research as main author.}
\affiliation{
  \institution{Hamburg University of Technology}
  \city{Hamburg}
  \country{Germany}
  }
\email{nicolas.diaz-ferreyra@tuhh.de}

\author{Gautam Kishore Shahi}
\authornotemark[1]
\affiliation{%
  \institution{University of Duisburg–Essen}
  \city{Duisburg}
  \country{Germany}
  }
\email{gautam.shahi@uni-due.de}

\author{Catherine Tony}
\affiliation{
  \institution{Hamburg University of Technology}
  \city{Hamburg}
  \country{Germany}
  }
\email{catherine.tony@tuhh.de}

\author{Stefan Stieglitz}
\affiliation{%
  \institution{University of Potsdam}
  \city{Potsdam}
  \country{Germany}
  }
\email{stefan.stieglitz@uni-potsdam.de}

\author{Riccardo Scandariato}
\affiliation{
  \institution{Hamburg University of Technology}
  \city{Hamburg}
  \country{Germany}
  }
\email{riccardo.scandariato@tuhh.de}

\renewcommand{\shortauthors}{D\'{i}az Ferreyra et al.}

\begin{abstract}

During the outbreak of the COVID-19 pandemic, many people shared their symptoms across Online Social Networks (OSNs) like Twitter, hoping for others' advice or moral support. Prior studies have shown that those who disclose health-related information across OSNs often tend to regret it and delete their publications afterwards. Hence, deleted posts containing sensitive data can be seen as manifestations of online regrets. In this work, we present an analysis of deleted content on Twitter during the outbreak of the COVID-19 pandemic. For this, we collected more than 3.67 million tweets describing COVID-19 symptoms (e.g., fever, cough, and fatigue) posted between January and April 2020. We observed that around 24\% of the tweets containing personal pronouns were deleted either by their authors or by the platform after one year. As a practical application of the resulting dataset, we explored its suitability for the automatic classification of regrettable content on Twitter. 
\end{abstract}




\maketitle

\section{Introduction}
\label{sec:1}

Online Social Networks (OSNs) have become mainstream communication channels central to our daily lives. Platforms like Twitter or Instagram are frequented daily by millions of users who create, share, and engage with media content ranging from (fake) news to memes \cite{shahi2022amused}. The disclosure of personal information in OSNs is a common practice and a necessary condition for establishing and strengthening interpersonal relationships with others \cite{towner2022revealing}. During the COVID-19 pandemic, online self-disclosure became a central instrument for social interaction as lockdowns prevented people from leaving their homes and participating in onsite gatherings \cite{towner2022revealing}. Particularly, many were compelled to interact with their peers almost exclusively through OSNs in order to satisfy their social, informational, and entertainment needs \cite{westin2022just}. In turn, the exchange of (personal) information via social media channels became critical for people's connectedness and psychological well-being in light of the contact restrictions imposed worldwide \cite{nabity2020inside}.

As the COVID-19 pandemic unfolded, the healthcare sector became rapidly saturated with new patients and had to start prioritising efforts \cite{westin2022just,nabity2020inside}. Hence, the number of appointments in clinics and hospitals was limited and granted only to those at a higher risk or showcasing severe symptoms. Given the impossibility of obtaining fast medical assistance, many turned to OSNs in search of answers about the manifestations, treatments, and side effects of a disease that, until then, was unknown to the whole world \cite{haggag2021covid,bhatt2022user}. The result was a flood of sensitive, health-related information disseminated across the Internet in exchange for guidance, instrumental assistance, and social support from others \cite{bhatt2022user,zhang2021distress}. Nevertheless, disclosing such information in mainstream social media channels can lead to privacy threats, including stigma, discrimination by insurance companies, and public judgement \cite{Blose_2021,bridges_2021,liao2019sharing}. Moreover, it may diminish the amount of social support from peers and even derive in regrettable online experiences \cite{zhang2021distress}.

\subsubsection*{\textbf{Motivation}}

Regrets can be seen as negative emotions caused by faults in one's actions \cite{guha2018regrets}. Such emotions come along with a high load of self-blame, as those who experienced regrets often wonder, \textit{``what would have happened?''} if they had made a different decision \cite{renaud2022regrets}. Online regrets are frequently caused by a mismanagement of audiences and personal information in OSNs. Particularly, they are often the outcome of users' flawed estimations about the size and composition of the group of individuals targeted by their publications \cite{guha2018regrets}. Prior research has shown that regrettable experiences can have a large impact on the way people engage with and through social media channels. For instance, those who regret having shared a post with personal information are prone to delete it afterwards and even change their self-disclosure behaviour \cite{wang2011a}. Hence, individuals can, to a certain extent, learn from these (negative) experiences and leverage them for their own psychological growth \cite{saffrey2008praise}. Still, the body of knowledge on online regrets remains scarce and demands further research exploring their role in people's cybersecurity decisions  \cite{towner2022revealing,renaud2022regrets}.

In a moment of crisis and confusion, making rational privacy decisions is even more challenging as they can be impaired by users' urgency and sense of emergency \cite{bol2022skype,zhang2021distress, umar2021self}. Hence, they often translate into large sets of posts that, given their sensitivity, may be regretted later on by their authors \cite{renaud2022regrets}. The COVID-19 pandemic has motivated many research efforts seeking to understand people's online self-disclosure decisions during a crisis of such dimensions \cite{squicciarini2020tipping}. Such efforts have resulted in several methods (and data sets) characterising the dissemination of personal information across OSN platforms \cite{bhatt2022user,umar2021self} and empirical studies zooming into users' privacy concerns during this period \cite{westin2022just,haggag2021covid,zhang2021distress}. Nevertheless, little attention has been placed on regrettable self-disclosure practices during the pandemic and their negative effects on OSN users \cite{towner2022revealing}. Furthermore, to the best of our knowledge, the use of actionable information on regrets has not been extensively explored for the elaboration of preventative technologies (e.g., nudges and awareness mechanisms alike) nor thoroughly documented within the current literature.

\subsubsection*{\textbf{Contribution and Research Questions}}

Prior research in online self-disclosure has emphasised the catalyst role of regrets for users' \textit{self-cleaning} practices in OSNs \cite{bhattacharya2016characterizing,zhou2016tweet,wang2011a}. That is, regarding their influence on people's decision to revise and delete posts containing personal and sensitive information. Based on this premise, this work elaborates on regrettable self-disclosure practices on Twitter during the outbreak of the COVID-19 pandemic. Particularly, it introduces an approach for the identification of deleted posts containing health-related information, while exploring their suitability for the automatic classification of online regrets. For this, we posit and address the following research questions (RQs): 
\begin{itemize}
    \item \textbf{RQ1}: \textit{How can a corpus of regrettable, health-related COVID-19 tweets be generated?} To answer this RQ, we conducted a search of COVID-related tweets published during the outbreak of the pandemic. For this task, we focused primarily on tweets describing COVID-19 symptoms (e.g., fever, cough, and fatigue) and including personal pronouns (e.g., `I', `me', and `she'). Then, using Twitter's API, we checked their publication status after one year and labelled them as ``deleted'' or ``not deleted'', accordingly. Overall, our resulting dataset contains 17.213 deleted tweets corresponding to the period January-April 2020.
    \item \textbf{RQ2}: \textit{Can such a corpus be leveraged for the automatic identification of online regrets?} We addressed this RQ by training different Machine Learning (ML) and Deep Learning (DL) models using the resulting dataset. Particularly, we assessed the performance of state-of-the-art ML algorithms along with mainstream DL methods, including Support Vector Machine (SVM), Long Short-Term Memory (LSTM) and Bidirectional Encoder Representations from Transformers (BERT). This last one yielded the best performance in terms of F1 and recall scores, followed by Naive Bayes when it comes to precision.
\end{itemize}

These findings contribute to understanding and characterising the interplay between online regrets and self-cleaning practices in OSNs, particularly under situations of crisis like the one unleashed by the COVID-19 pandemic. Moreover, they provide actionable information and empirical evidence on such interplay while showcasing its relevance for developing user-centred privacy solutions.

%\textit{Paper Structure:} Avoid if not much space left ...

\section{Background and Related Work} \label{sec:background}

%Large-scale crises like the COVID-19 pandemic have been matter of privacy research in the late years. Especially, with regard to people's online self-disclosure behaviour and privacy concerns within Computer-Mediated Communication~(CMC) settings. In the following sections we discuss some of the main findings documented throughout the current literature and provide the necessary background information for the approach presented in this paper.

%The COVID-19 pandemic has been matter of concern and 

\subsubsection*{\textbf{Online Self-Disclosure During Crises}}

People's online self-disclosure behaviour has attracted researchers from different disciplines thanks to the disruptive emergence of OSNs in the late 2000s. Since then, several works in CMC, media and cognitive psychology have sought to understand the motivations, benefits, and drawbacks behind this multifaceted process \cite{kramer2019mastering}. The \textit{privacy paradox} and the \textit{privacy calculus} have emerged from these efforts as the two most compelling theories characterising the self-disclosure practices of OSN users \cite{kezer2022getting,kramer2019mastering}. Whereas the former describes an offset phenomenon between people's expressions of privacy concerns and their ultimate behaviour, the latter suggests that such behaviour is closely related to people's ability to anticipate the risks and benefits of their self-disclosure acts \cite{princi2020out}. Nevertheless, recent work has challenged the postulates of the privacy calculus, suggesting that people's privacy decisions are guided mainly by heuristic judgements instead of rational risk estimations \cite{ostendorf2022,sundar2020online,waldman2020cognitive}. For instance, users of OSNs tend to use the self-disclosure practices of their trusted peers as a reference for what is adequate or not to post \cite{marmion2017cognitive}. Moreover, they are prone to reveal more personal information in their posts if they perceive the platform as trustworthy and credible. All in all, heuristics reduce the cognitive effort demanded by complex self-disclosure decisions entailing significant long-term consequences \cite{marmion2017cognitive}. Still, they can lead to sub-optimal privacy choices and, consequently, to unwanted incidents such as reputation damage, scamming, or identity theft \cite{ostendorf2022theoretical}.

Alongside, prior research has also delved into people's self-disclosure practices during crises and extreme events. \citet{Jung2012a} was one of the first to address this matter with a survey among college students after the 2011 earthquake in Japan. The findings of this study indicate that, during such a catastrophic event, people turned to OSN channels to inform others that they were safe, to check on family and friends, and share their own impressions of the earthquake (e.g., through videos and photos). Subsequent related studies have consistently shown that such information, along with individuals' self-reported location, is of great value for public health agencies to deploy timely and efficient first-aid strategies \cite{muniz2020social}. Regarding the COVID-19 pandemic, \citet{westin2022just} found that people spent more time online as a consequence of health restrictions and lockdowns imposed during the outbreak. Particularly, they felt pressured to stay ``relevant'' online by posting more about their thoughts and activities as well as re-sharing old content. \citet{Blose_2021} also observed an increased number of Twitter publications portraying users' emotional states and personal experiences during this period. Moreover, \citet{umar2021self} showed that self-disclosure levels remained high even months after the World Health Organisation's pandemic declaration on March 11, 2020. 

\subsubsection*{\textbf{Online Regrets and Self-Cleaning}}

Although the benefits and drawbacks of online self-disclosure have been extensively documented within the current literature, the body of knowledge on regrets remains relatively scarce \cite{towner2022revealing}. The work of \citet{wang2011a} is perhaps one of the first ones addressing people's regrettable experiences in OSNs through an empirical study among Facebook users. Such a study revealed that those who had suffered a negative self-disclosure experience tended to delete their publications afterwards. This strategy, often referred to as \textit{self-cleaning}, was later investigated by \citet{almuhimedi2013tweets} over a corpus of 1.6 million deleted tweets. Among their findings, they observed a higher frequency of negative words in deleted tweets when compared to non-deleted ones, but also a substantial amount of tweets deleted due to superficial reasons (e.g., typos). Alongside, a study by \citet{xie2015see} with adolescents showed that regrets are more likely to occur among those who befriend strangers, have a large network, and use OSNs regularly. Moreover, prior research has reported that deleted publications often portray sensitive, health-related information \cite{geetha2019tweedle}, and that regrettable self-disclosure experiences can lead people to deactivate their OSN accounts or even withdraw entirely from social media \cite{guha2018regrets}.

Prior investigations have also addressed the elaboration of Machine Learning (ML) and Deep Learning (DL) models for the automatic detection of regrettable OSN publications. For instance, \citet{petrovic2018a} used Support Vector Machines (SVMs) and Passive Aggressive (PA) algorithms to predict deleted content on Twitter. These models were trained using \textit{social}, \textit{author}, and \textit{text} features (e.g., hashtags and number of followers), yielding an F1 score of 0.27 at its best. Further work has sought to improve the performance of deleted tweet classifiers by leveraging different ML/DL algorithms and model features. Such is the case of \citet{bagdouri2015a}, who explored the role of tweet-based and user-based features for predicting the deletion of tweets composed in Arabic. Likewise, \citet{zhou2016a} trained a Naive Bayes classifier employing users' historical deletion patterns as a model feature, whereas \citet{gazizullina2019a} applied unsupervised learning to optimise the selection of such features. Overall, these models have yielded a significantly higher performance when compared to other state-of-the-art approaches. Yet, up to our knowledge, empirical evidence on self-cleaning practices during global crises like the COVID-19 pandemic has not been documented in the current literature. Moreover, the automatic classification of deleted tweets in this context remains unexplored and calls for further investigations.

\section{Methodology}
\label{sec:4}

To address the RQs formulated in this paper, we collected a set of tweets published during the outbreak of the COVID-19 pandemic. We focused on publications portraying health-related issues, as they are more likely to be regretted by their authors in the future \cite{geetha2019tweedle}. Given the interplay existing between online regrets and deleted content in OSNs, we also checked whether these tweets were still available after one year. The resulting dataset, named \textit{RegretCovid}, was then assessed for the automatic classification of regrettable publications. The following sections describe the methodology we applied for the corresponding data collection and classification tasks.

\subsection{Data Collection} 

% \begin{figure}[!t]
% \includegraphics[width=0.9\textwidth]{graphics/method.png}
% \caption{Methodology applied for the curation of the ``RegretCovid'' dataset.}
% \label{fig:data_collection}
% \end{figure}

The upper part of \autoref{fig:merged_method} illustrates the steps we followed for curating the ``RegretCovid'' dataset:

\begin{enumerate}[leftmargin=*]
    \item \textbf{Tweets retrieval:} We used Tweepy \cite{shahi2021exploratory, shahi2022shapes}, a Twitter crawler written in Python, to mine an initial set of health-related tweets. Such tweets were collected by executing several search requests with trending hashtags and keywords such as \#fever, \#cough, ``shortness of breath,'' \#sore, \#throat, and \#fatigue. Overall, we retrieved 8.9 million tweets written in English and published between January 27 and April 30, 2020. Since our main goal was to investigate online self-disclosure during the pandemic, we further filtered the tweets containing personal pronouns (e.g., I, me, you, and she) using the TextBlob library \cite{loria2022textblob}. This resulted in a dataset of approximately 3.67 million tweets deemed as ``personal''.
    \item \textbf{Extraction of COVID-19 related words}: Since our hashtag-guided search could lead to instances that are not relevant to our study (e.g., tweets describing general flu symptoms), we applied an additional filter based on frequent Covid-related terms. For this, we computed the frequency of each unique term in the ``Personal tweets'' dataset and generated a list with those having a frequency higher than 100. After a manual inspection, we narrowed down such a list to 58 Covid-related words, including `covid-19', `sars-cov-2', and `wuhanvirus'. 
    \item \textbf{Filter personal COVID-19 tweets}: We used the list obtained in the previous step to build a set of Covid-related personal tweets. That is, by filtering out those tweets that did not contain any of the 58 terms included in the list. After removing duplicates, we ended up with a corpus of 886.145 personal COVID-19 tweets.
    \item \textbf{Label deleted tweets}: Using Tweepy, we checked whether these tweets had been deleted after one year. This was done using the \texttt{get\_status} method included in the library and setting 01-07-2021 as the cutoff date (i.e., exactly one year after the initial Tweets retrieval). All in all, we determined that 241.307 tweets had been deleted by then, whereas 671.838 remained accessible. We applied a final refinement on this dataset by removing retweets, duplicates and tweets from suspended accounts. This resulted in the final ``RegretCovid'' dataset comprising 17.213 deleted and 122.083 non-deleted COVID-19 tweets.
\end{enumerate}

A chronological summary of the datasets generated throughout the method steps is presented in \autoref{tab:dataset_summary}. %A temporal evolution of deleted and non-deleted tweets is depicted in \autoref{fig:timeline}.

\begin{table*}[!t]
  \centering
   \caption{Chronological summary of the tweets collected and processed throughout the curation of the \textit{RegretCovid} dataset.}
  \begin{tabular}{ l c c c c c} 
\toprule
\textbf{Tweets} & \textbf{January} & \textbf{February} & \textbf{March} & \textbf{April} & \textbf{TOTAL}  \\ \midrule
Health-related tweets & 109.029 & 775.663 & 1.361.453 & 3.318.594 & 8.883.333 \\
Personal health-related tweets & 52.024 & 407.330 & 799.185 & 2.411.723 & 3.670.262 \\
Personal COVID-19 tweets & 6.606 & 21.130 & 716.610 & 138.510  & 886.145 \\
Deleted tweets (*) & 1.599 (188) & 9.018 (764) & 207.508 (10.983) & 38.214 (5.183) & 214.307 (17.213) \\
Non-deleted tweets (*) & 5.007 (1.530) & 12.112 (5.539) & 509.102 (78.574) & 100.256 (35.485) & 671.838 (122.083) \\
\bottomrule
\multicolumn{6}{l}{\underline{Note}: (*) Tweets included in the final dataset are shown between parenthesis.}
  \end{tabular}
  \label{tab:dataset_summary}
\end{table*} 

% \begin{figure}
% \includegraphics[width=1\textwidth]{graphics/tweets_plot.jpg}
% \caption{Timeline plot of both Deleted and non-deleted tweets}
% \label{fig:timeline}
% \end{figure}

\subsection{Classification of Regrettable Tweets}

\begin{figure*}[!t]
\includegraphics[width=0.7\textwidth]{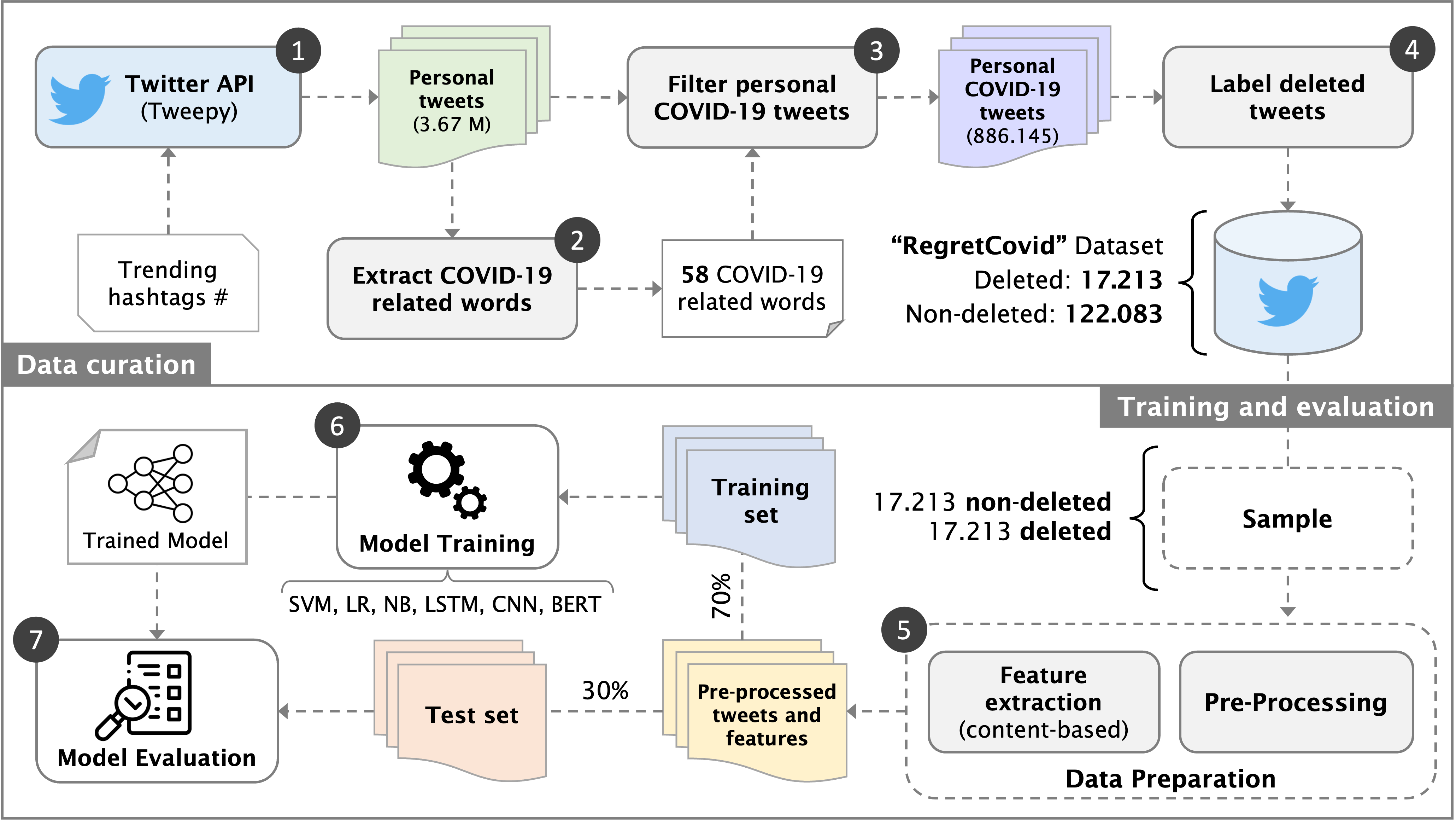}
\caption{Methodology applied for (i) curating the ``RegretCovid'' dataset, and (ii) training the ML/DL models.}
\label{fig:merged_method}
\end{figure*}

The lower part of \autoref{fig:merged_method} illustrates the methodology employed for training a deleted tweets classifier using different ML/DL models. Similar to previous work (Section~\ref{sec:background}), we used Naive Bayes (NB), Logistic Regression (LR) and Support Vector Machines (SVMs) for ML along with Long Short-Term Memory (LSTM), Convolutional Neural Networks (CNNs) and the Bidirectional Encoder Representations from Transformers (BERT) for DL. As shown in \autoref{tab:dataset_summary}, there is a high-class imbalance in our dataset since the number of non-deleted tweets is higher than the number of deleted ones. Hence, we first took a random sample of 17.213 non-deleted tweets (all containing 6 words or more) and proceeded as follows:
\begin{enumerate}[leftmargin=*] \setcounter{enumi}{4}
    \item \textbf{Data Preparation} 
    \begin{itemize}
        \item \textit{Pre-processing}: Prior to model training, we pre-processed the tweets in the sampled \textit{RegretCovid} corpus following standard techniques for Natural Language Processing (NLP) \cite{Egger2022, shahi2020fakecovid,nandini2022explaining}. This included removing punctuation, e-mails, URLs, and short words (i.e., of less than 3 characters) along with lowercasing. Thereby, we sought to remove noise from the data that may negatively impact the overall computational cost and performance of the trained models.
        \item \textit{Feature extraction}: We used the Term Frequency - Inverse Document Frequency (TF-IDF) for feature extraction in all ML models (i.e., SVM, LR, and NB). For this, we used the Sklearn TF-IDF transformer\footnote{\url{https://scikit-learn.org/stable/modules/generated/sklearn.feature_extraction.text.TfidfTransformer.html} [Accessed: 02.03.2023]} with its default configuration. For LSTM, we applied Text-to-Sequences to vectorise the tweets, whereas for CNN we applied a Document-Term Matrix. In the case of BERT, we chose Hugging Face's pre-trained \texttt{distilbert-base-uncased} model\footnote{\url{https://huggingface.co/distilbert-base-uncased} [Accessed: 02.03.2023]}, which uses WordPiece as tokenizer.
    \end{itemize}
    \item \textbf{Model Training}: For all models, we used 70\% of the corpus for training and the remaining 30\% for testing. Grid search, in combination with k-fold cross-validation, was used to optimise the hyperparameters of the SVM and LR models. For the former, we adopted C=0.1 and an RBF kernel, whereas for the latter we used C=1 with L1 regularisation. After running cross-validation with different fold values (3 to 7), we selected k=6 as it produced the best performance in terms of variance and bias. In the CNN classifier, we chose a four-layer model with a kernel size of 4 together with a cross-entropy loss function and an Adam optimiser. For BERT's fine-tuning, we used a learning rate of 2e-5, batch size 16, and a sequence length of 128. In the case of LSTM we adopted a batch size of 8, whereas for CNN we chose a four-layer model with a kernel size of 4 together with a cross-entropy loss function and an Adam optimiser. 
    \item \textbf{Model evaluation}: We evaluated the performance of the generated model in terms of \textit{precision} (i.e., percentage of tweets correctly classified as ``regrettable''), \textit{recall} (i.e., ratio of deleted tweets classified as such), and \textit{F1} (i.e., the harmonic mean between precision and recall). Results are summarized in \autoref{tab:results} and discussed in Section~\ref{sec:results}.
\end{enumerate}

% \begin{figure}
% \includegraphics[width=0.9\textwidth]{graphics/covid_training.png}
% \caption{Methodology employed for training and evaluating the ML/DL models.}
% \label{fig:classification}
% \end{figure}

\subsection{FAIR Data and Ethical Considerations} \label{sec:ethics}

We have conducted all experiments on a macro level following strict data access, storage, and auditing procedures for the sake of accountability. The data collection and methodology used in this paper were approved by the ethical committee of the University of Duisburg-Essen, Germany with project code 2209PKSG8058. The \textit{RegretCovid} dataset, together with the ML/DL models generated in this study, are available in the paper's \textbf{Replication Package}\footnote{\url{https://doi.org/10.5281/zenodo.7692521}}. We provide the ID of all collected tweets and a Python script to download them.

%All tweets in the dataset have been anonymised by removing personally identifiable information (e.g., profile details, IDs).

\section{Results and Discussion}
\label{sec:results}

\subsubsection*{\textbf{RegretCovid dataset (RQ1)}}

From \autoref{tab:dataset_summary}, we can determine that just around 10\% of the collected tweets contain personal health-related information about the COVID-19 pandemic (i.e., 886.145 out of 8.88 million). We can also observe that about 24\% of these publications were deleted after one year either by their authors or by the platform (in this case, Twitter). However, when further analysing those publications that have been removed by the users themselves, we see that they just account for 1.94\% of the total (i.e., 17.213 out of 886.145). Notably, about 84\% correspond to retweets or posts that were published by a Twitter account that no longer exists. 

Although closing/deactivating OSN accounts has been deemed as a self-cleaning strategy (see \cite{guha2018regrets}), it was not possible for us to determine the specific reason why the accounts linked to our dataset were removed. That is, whether these accounts were removed by their respective owners or by the platform as a consequence of non-compliance with Twitter's terms of service \cite{twitterPolicy2022}. Consequently, we opted not to include these publications in our final corpus.

All in all, our results resemble the ones of prior work addressing self-disclosure during the COVID-19 pandemic. For instance, \citet{Blose_2021} conducted a similar analysis over a corpus of 53 million Covid-related tweets collected between January and August 2020. Their findings suggest that about 19\% of these tweets reveal personal details of the user who wrote them. A similar study by \citet{umar2021self} also revealed that around 28\% of the COVID-19 discussions within persistent groups of users elaborate on health-related topics. Regarding online regrets, \citet{geetha2019tweedle} observed around 60\% of regretted tweets contain health-related information about their authors. However, it should be noted that their study (unlike ours) did not analyse regrets through the lens of self-cleaning and was conducted over a dataset of publications unrelated to COVID-19. Hence, it is not surprising that these same proportions of regrettable tweets are not reflected in our corpus. In this sense, our findings are closer to the ones of \citet{almuhimedi2013tweets}, who observed a 2.4\% of deleted posts within a corpus of 1.6 million tweets.

\begin{table}
  \centering
   \caption{Classification Result of Self-disclosure tweets}
  \begin{tabular}{ c c c c c  } 
\toprule
\textbf{Model} & \textbf{Feature Extractor}  &  \textbf{Precision}   & \textbf{Recall}  &  \textbf{F1 Score}  \\
\midrule
SVM & TF-IDF &  0.57  &    0.56   &   0.56  \\ 
LR & TF-IDF &  0.56  &    0.56   &   0.56  \\ 
NB & TF-IDF &  \textbf{0.58}  &    0.56   &   0.55  \\ 
LSTM & Text2Sequence & 0.55   &   0.55   &   0.54  \\
CNN & Document Matrix & 0.54   &   0.54   &   0.54  \\
BERT & BERT uncased & 0.54 & \textbf{0.74} & \textbf{0.62} \\ \bottomrule
  \end{tabular}
  \label{tab:results}
\end{table}

\subsubsection*{\textbf{Model performance (RQ2)}} \autoref{tab:results} summarises the performance yielded by each ML/DL approach. We can see that there is no big difference in the precision of these models, which is 0.56 on average and 0.58 at its best (i.e., with an NB classifier). Regarding recall, BERT obtained the highest score of 0.74, whereas the rest of the models performed between 0.54 and 0.56. The same occurred in terms of F1, where the BERT classifier achieved the highest value of 0.62, while the rest yielded scores in the range of 0.54 and 0.56. These results outperform the ones from some previous investigations addressing the classification of deleted content on Twitter (e.g., \cite{bagdouri2015a,zhou2016a}). To the extent of our knowledge, higher recall and F1 values have only been achieved by \citet{gazizullina2019a} reaching 0.9 and 0.75, respectively. Still, these and other state-of-the-art approaches were built using generic Twitter datasets, whereas our ML/DL models have been trained using a Covid-specific corpus.

%For instance, \citet{bhatt2022user} curated a dataset of 114.296 tweets published between January and July 2020 discussing privacy issues steaming from the collection of health-related information by governments and other public agencies (e.g., via contact-tracing apps). 

%Presence of first person pronouns is generally considered as linguistic markers for self-disclosing texts [15]. Moreover, several studies have used presence of first person pronouns as a prominent feature in determining self-disclosure in online contents [2, 11, 33].

\subsubsection*{\textbf{Implications, recommendations, and limitations}} Informing users about potentially regrettable actions can contribute significantly to their well-being and shaping trustworthy OSN environments \cite{guha2018regrets,diazPST18}. Prior studies have pointed towards developing nudging solutions to support users' self-disclosure decisions on the Internet and mitigate their chances of experiencing regret. For instance, \citet{diaz2019umapadj} proposed a conceptual framework for learning regrettable self-disclosure patterns from deleted content on Facebook and used them to generate privacy-aware behavioural interventions. Particularly, they prescribe the use of such patterns for identifying and informing the users of OSNs about the presence of personal/sensitive data in their posts before publishing them.

%For instance, \citet{diaz2019umapadj} proposed a conceptual framework for learning regrettable self-disclosure patters from deleted content on Facebook. Such patterns could be used thereafter to elaborate behavioural interventions for nudging users' self-disclosure practices. That is, by informing them about the presence of personal/sensitive data in their posts before publishing them.

Our work contributes in this regard with a dataset of deleted COVID-19 tweets and ML/DL models that can be leveraged for the identification of potentially regrettable posts. Nevertheless, we acknowledge that there is room for improvement, especially when it comes to the performance of our classifier. Missing positive instances of ``potentially regrettable'' posts would be unacceptable within the intervention approach described above. However, false positives would be acceptable as nudges, at their core, are soft-paternalistic mechanisms that persuade (but not coerce) individuals. Hence, even if a message is misclassified, it should be the user who decides in the end whether to publish it or not. Therefore, when it comes to the performance of the ML/DL models, a higher recall score would be more critical than a higher precision.

Some limitations related to the strategy applied for building our dataset should be acknowledged and addressed in future publications. First, although personal pronouns are generally considered as a proxy for self-disclosing posts, other aspects such as feelings, degree of intimacy, and personal relationships should also be taken into account \cite{wang2016disclosure}. An unsupervised annotation method like the one proposed by \citet{umar2021self} could be employed to overcome the shortcomings of our current approach in this regard. Besides, narrowing down the search to first-person pronouns could also be a good alternative. Still, we opted for including third-person pronouns as well since privacy issues may also arise when users reveal personal details from others without their consent \cite{zhong2018multiparty}. Finally, the identification of regrets through the lens of deleted tweets should be further investigated with additional studies and annotation schemes (e.g., using ``Regret Theory'' as in \cite{geetha2019tweedle}) to avoid confounds.

Conducting research on deleted content exposes the limitations of current self-cleaning practices and raises some ethical questions. Particularly, it shows that more than deleting a tweet is needed to guarantee its complete removals from repositories outside Twitter (e.g., search engines and third-party applications) \cite{almuhimedi2013a}. Although enforcing deletion across repositories is hard (if not impossible), platforms should notify users about such a limitation and promote ethical data-mining principles among practitioners \cite{almuhimedi2013a}. Regarding this last point, we have consulted the ethical board of our university and taken the corresponding actions to minimise the privacy risks in the employed research methodology~(as described in \autoref{sec:ethics}).

\section{Conclusion and Future Work}
\label{sec:7}

Regrets (both anticipated and experienced) play a significant role in people's cybersecurity behaviour \cite{renaud2022regrets}. Moreover, although they entail negative feelings of sadness and remorse, regrets can be valuable resources for promoting individuals' psychological growth and self-awareness. In this work, we explored the interplay between regrets and self-cleaning practices on Twitter during the outbreak of the COVID-19 pandemic. The resulting dataset provides not only empirical evidence about such practices but also actionable information for the automatic identification of regrettable publications in OSNs. In that sense, the ML/DL models we trained show how regrets can be translated into concrete cybersecurity countermeasures seeking to aid users' online self-disclosure practices and protect the contextual integrity of their private information. Yet, further insights are needed to (i) improve the overall performance of such models and (ii) document the relationship between regrets and self-cleaning with greater detail. We will address these points in future investigations, along with the extrapolation of our findings to other application domains (i.e., beyond COVID-19). We also plan to assess the shortcomings of the generated models by conducting an error analysis (c.f., \cite{rochert2021networked, nandini2022explaining}). That is, a deep inspection of misclassified test instances to understand which characteristics of the models (e.g., features, training data) contribute (or not) to their overall performance.

\bibliographystyle{ACM-Reference-Format}
\bibliography{references}

\end{document}